\documentclass[aps,pre,twocolumn]{revtex4-1}
\usepackage{amsmath,amssymb,graphicx}
\input epsf
\usepackage{amsbsy}
\usepackage{latexsym}
\usepackage{color}
\usepackage{graphicx}
\usepackage{psfrag}
\usepackage[normalem]{ulem}
\usepackage{float}
\usepackage{dsfont}

\usepackage{booktabs}

\newcommand{\be}{\begin{equation}}
\newcommand{\ee}{\end{equation}}

\begin{document}

\title{Stability of synchronous states in sparse neuronal networks}

\author{Afifurrahman, Ekkehard Ullner, Antonio Politi}
\affiliation{Institute for Pure and Applied Mathematics and Department of Physics (SUPA), Old Aberdeen, Aberdeen AB24 3UE, United Kingdom}
 
\begin{abstract}
The stability of synchronous states is analysed in the context of two populations of inhibitory and excitatory neurons, characterized by different pulse-widths. The problem is reduced to that of determining the eigenvalues of a suitable class of sparse random matrices, randomness being a consequence of the network structure. A detailed analysis, which includes also the study of finite-amplitude perturbations, is performed in the limit of narrow pulses, finding that the stability depends crucially on the relative pulse-width. This has implications for the overall property of the asynchronous (balanced) regime.
\end{abstract}

\maketitle

\section{Introduction}\label{sec:intro}

Networks of oscillators are widely studied in many fields: 
mechanical engineering~\cite{Blekhman_1995applmechrev,Blekhman_1997syscontlet}, power grids~\cite{filatrella_2008epjb},
arrays of Josephson junctions~\cite{hadley_baesley_1987apl}, cold atoms~\cite{javaloyes_perrin_politi_2008pre},
neural networks~\cite{Izhikevich}, and so on.
Such networks can be classified according to the single-unit dynamics, the coupling mechanism, the presence of heterogeneity, 
and network topology.
Since phases are the most sensitive variables to any kind of perturbation~\cite{pikovsky_rosenblum_kurths_2001},
most of the attention is devoted to setups composed of phase oscillators~\cite{winfree}, i.e. one-dimensional dynamical systems.
However, even the study of such relatively simple models is not as staightforward as it might appear. 
In fact, a wide variety of dynamical regimes can emerge even in mean field models of identical oscillators, 
ranging from full synchrony to splay states, and including hybrides, 
such as partial synchronisation~\cite{partial}, chimera~\cite{chimera}, and cluster states~\cite{Golomb92}.
General theory of synchronisation is, therefore, a much investigated field.

In this paper we focus on synchronous states by referring to a rather popular class of neural networks, but the whole formalism
can be easily extended to more general systems so long as the coupling is mediated by the emission of pulses.
In neuroscience the neuron dynamics is often described by a single variable, the membrane potential, which evolves according
to a suitable velocity field. The resulting model is equivalent to a phase oscillator, where the variable of the bare system increases
linearly in time while the complexity of the evolution rule is encoded in the phase response curve (PRC), which accounts for
the mutual coupling~\cite{Abbott93}.
Under the additional approximation of a weak coupling strength, the model can be further simplified and cast into a
Kuramoto-Daido form, where the coupling depends on phase differences between pairs of oscillators~\cite{Golomb01,PolitiRosenblum}.
Here, however, we stick to pulse-coupled oscillators. 

The stability of the synchronised state of pulse-coupled phase oscillators has been first studied in the context of 
excitatory $\delta$-pulses~\cite{Mirollo90}. Synchronisation is induced when two oscillators are sufficiently close and a 
common excitatory $\delta$-pulse instantaneously sets both to the same value. 
Later, the stability analysis for excitatory and inhibitory pulses~\cite{vanVreeswijk94,Hansel95} has been extended to $\delta$-pulses 
with continuous PRCs \cite{Goel02}.
General formulas are mostly available under severe restrictions, such as identical oscillators, mean field interactions,
or $\delta$-like pulses.

The $\delta$-like pulse assumption is particularly limiting, not only because realistic systems are characterized by a finite width, 
but also because it has been shown that zero-pulsewidth is a singular limit, which does not commute with the 
thermodynamic limit (infinitely large networks) -- at least in the context of splay states~\cite{Zillmer07}.  
Relaxing the zero-width limit forces to increase the phase-space dimension to account for the dynamics of the fields
felt by the different neurons. 
The most general result we are aware of is a formula derived in Ref.~\cite{Olmi14} for a single population of
identical neurons in the presence of mean-field coupling and the so-called $\alpha$-pulses. 

The introduction of sparseness implies a significant increase in the computational complexity because of the randomness of
the connections. In this context, the most relevant results are those derived in Ref.~\cite{timme_wolf_2008nonlin}, where
a sparse random network (Erd\"os-R\'enyi type) has been investigated in the presence of $\delta$-pulses. 
The approach is rather complex since the noncommutativity associated with changes in the order of the incoming spikes 
obliged the authors to introduce a multitude of linear operators to solve the problem.

Here, we extend this kind of stability analysis to finite pulse-widths in two populations of excitatory, respectively inhibitory,
neurons. Our approach can also be considered as an extension to sparse networks of the work in Ref.~\cite{Olmi14} devoted to
mean-field models.
This setup is chosen in studies of the so-called balanced state \cite{Brunel00}, where the asynchronous 
regime is dominated by strong fluctuations. Typically, the balance depends on both the relative size of the two populations
and the relative amplitude of the pulses. In this paper, a careful study of the fully synchronous regime shows
that also the relative pulse-width plays a non-trivial role.

Finite-width pulses can obviously have infinitely many different shapes. In this paper we consider the simplest case of 
exponential spikes and assume, as usual, that they superpose linearly.
In practice, this means that each oscillator (neuron) is characterized by three variables: the phase or, equivalently,
the membrane potential and two variables describing the incoming excitatory and inhibitory fields, respectively.
At variance with Ref.~\cite{Olmi14}, instead of transforming the model into a mapping (from one to the next
spike emission), here we preserve the time continuity, as this approach allows for a more homogeneous treatment of
the oscillators maintaining the full $3N$ dimensional structure of the phase-space (where $N$ is the number of oscillators).

Furthermore, in agreement with previous publications \cite{ostojic_2014nat,ullner_politi_torcini_2018chaos,politi_ullner_torcini_2018epjst} 
we assume that each neuron receives exactly the same number of excitatory and inhibitory synaptic connections. 
In fact, in spite of the random connectivity, in the fully 
synchronous regime, all neurons are characterized by exactly the same input. 
The degeneracy of the Lyapunov spectrum observed in mean-field models is lifted
and the stability must be assessed by determining the eigenvalues of a suitable (sparse) random matrix.

More precisely, in Sec.~\ref{sec:model} we define the model, including the specific phase response curve used to perform numerical tests.
The overall linear stability analysis is discussed in Sec.~\ref{sec:linstab}, first with reference to the general case and then specifically
referring to short (but finite) pulses. In the same section we also determine the conditional Lyapunov exponent $\lambda_c$, (i.e the
exponent describing the response of a single neuron subject to a given - periodic - forcing): at variance with the mean-field model, $\lambda_c$ differs
from the maximum exponent of the whole network, indirectly confirming the nontrivial role played by the connectivity.
In Sec.~\ref{sec:app}, we implement the formulas determined in the previous section to discuss the qualitative changes
observed by varying the relative pulse-width.
Finally, Sec.~\ref{sec:conclusion} is devoted to a summary of the main results and an outline of the open problems.

\section{Model}\label{sec:model}

The object of study is a network of $N$ phase-oscillators (also referred to as neurons), the first $N_e$ 
being excitatory, the last $N_i$ inhibitory (obviously, $N_e+N_i = N$).
Each neuron is characterized by the phase-like variable $\Phi^j\le 1$ (formally equivalent to the membrane 
potential), while the (directed) synaptic connections are represented  by the connectivity matrix $\mathbf{G}$ with the entries
\begin{align*}
& G_{j,k}=
    \begin{cases}
    1, & \text{if $k \to j$ active} \\
    0, & \text{otherwise } 
    \end{cases}
\end{align*}
where $\sum_{k\ge 1}^{N_e} G_{j,k} = K_e$ and
$\sum_{k > N_e}^{N} G_{j,k} =K_i$, meaning that each neuron $j$ is characterized by the same number of incoming
excitatory and inhibitory connections, as customary assumed in the literature \cite{ostojic_2014nat} 
($K=K_e+K_i$, finally represents the connectivity altogether).

The evolution of the phase of both excitatory and inhibitory neurons  is ruled by the same equation, 
\begin{equation}
   \dot{\Phi}^j  =   1 + J \, \Gamma\left(\Phi^j\right)\left(E^j - I^j\right)\label{eq:mod1} \, ,
\end{equation}
where $\Gamma(\Phi)$ represents the phase-response curve (PRC), $J$ the coupling strength and $E^j$ ($I^j$) the excitatory (inhibitory) 
field generated by the incoming connections. He we assume that $K$ as well as $J$ are independent of $N$, i.e. we refer to
sparse networks.
Whenever $\Phi^j$ reaches the threshold $\Phi_{th}=1$ , the phase is reset to $\Phi_r = 0$ and 
enters a refractory period $t_{r}$ during which it stands still and is insensitive to the action of the
excitatory ($E^j$) and inhibitory ($I^j$) field. At the same time, the fields of the receiving neurons are activated. 
If the neuron $k$, emitting a spike at time $t^k_n$, is excitatory ($k \le N_e$), then the excitatory field $E^j$ of any receiving 
neuron $j$ is activated (and similarly for the inhibitory field $I^j$).

The fields in Eq.~(\ref{eq:mod1}) evolve according to the differential equations
\begin{eqnarray}
\dot{E}^j    &=& -\alpha\left( E^j - \sum_{n}  G_{j,k} P_{k,k} \delta(t-t^k_n) \right)  \label{eq:mod2} \\
\dot{I}^j    &=& -\beta \left( I^j - g\sum_{n} G_{j,k} (\delta_{k,k}-P_{k,k})\delta(t-t^k_n) \right) \, ,\nonumber
\end{eqnarray}
where $\alpha$ ($\beta$) denotes the inverse pulse-width of the excitatory (inhibitory) spikes.
The coefficient $g$ accounts for the relative amplitude of inhibitory spikes compared to excitatory ones.
$P_{k,m}$ represents the elements of a projector operator $\mathbf{P}$, separating excitatory from inhibitory neurons:
$P_{k,m} = 0$ except when $k=m \le N_e$, in which case $P_{k,k} = 1$. 

\begin{figure}
\center{\includegraphics[width=0.43\textwidth]{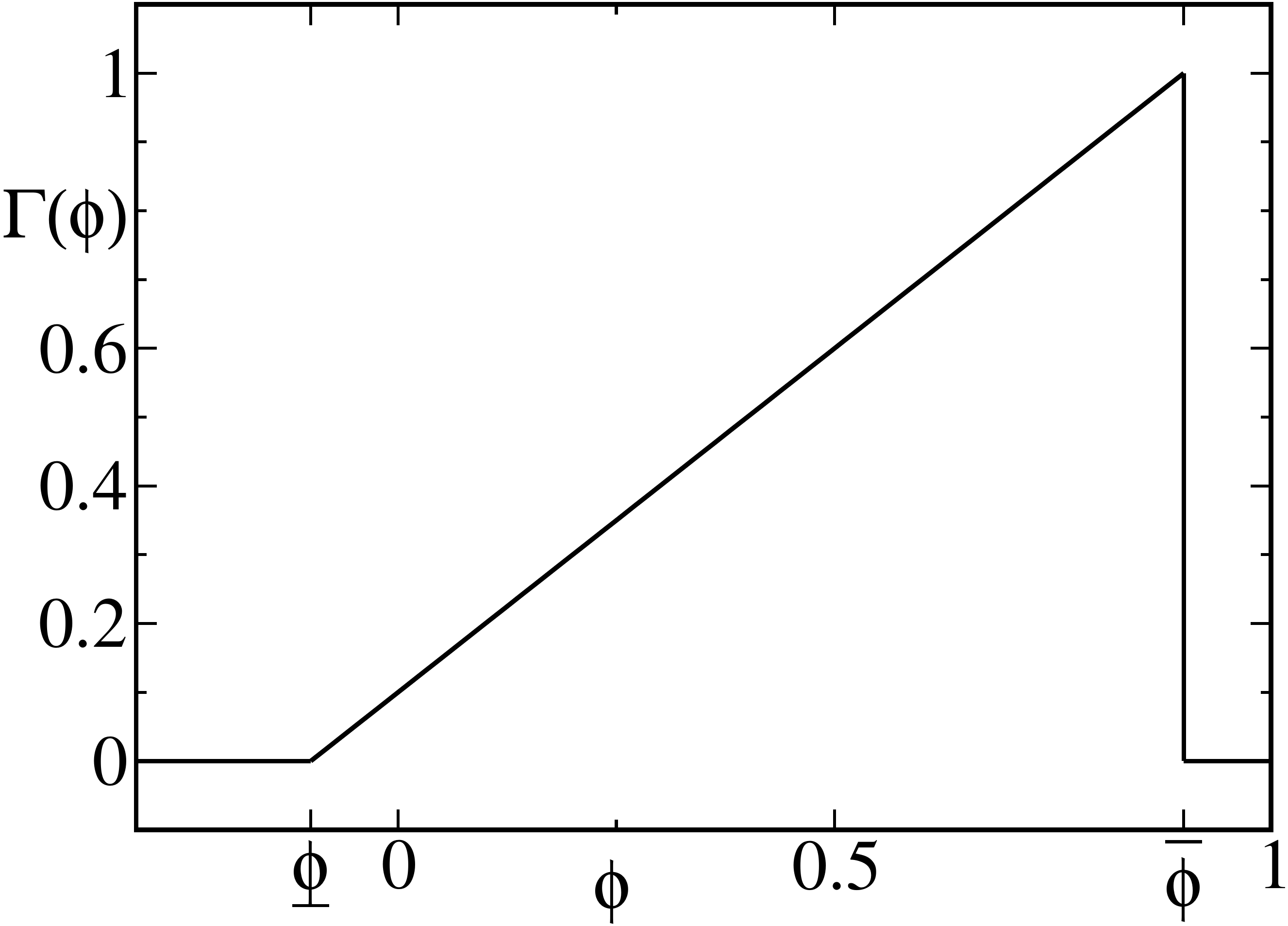}}
\caption{Example of the phase response curve (PRC) used in Sec.~\ref{sec:app} with $\underline \Phi = -0.1$, $\overline \Phi = 0.9$ in combination with $\Phi_{r}=0$ and $\Phi_{th}=1$.}
\label{fig:PRC}
\end{figure}

In order to be more specific, we introduce the PRC used later on as a testbed for the formalism developed in
the next section. We have chosen to work with the following piecewise linear PRC,
\begin{equation}\label{eq:PRC}
  \Gamma(\Phi^j) =
  \begin{cases}
  \left(\Phi^j -\underline{\Phi}\right) & \text{if $\underline \Phi<\Phi^j <\overline \Phi$} \\
  0  & \text{otherwise}
  \end{cases}
\end{equation}
where $\underline \Phi<0$, and $0<\overline \Phi<1$ characterize the PRC.
The resulting shape is plotted in Fig.~\ref{fig:PRC} for $\underline \Phi = -0.1$ and 
$\overline \Phi = 0.9$~\footnote{The peculiarity of a vanishing PRC for $\overline \Phi < \Phi \le 1$
is introduced to match the PRC recently introduced in~\cite{ullner_politi_2006prx} to mimic delay-coupled leaky 
integrate-and-fire neuron. However, it does not affect the generality of the formalism, as it can be easily
``removed'' by assuming $\overline \Phi = 1$.} .

As anticipated in the introduction, we are interested in assessing the stability of the fully synchronous dynamics of period $T$ as
a function of the relative pulse-width, where $T$ is the interspike interval. The solution is obtained by integrating the equation,
\begin{align}\label{eq:sync}
\begin{cases}
\dot\Phi = {1} + J \, \Gamma(\Phi)(E-I), \\
E(t)=E_{\circ}\mathrm{e}^{-\alpha t} \\
I(t)=I_{\circ}\mathrm{e}^{-\beta t}, 
\end{cases}   
\end{align}
where
$$
E_\circ = \frac{K_e\alpha}{1-\mathrm{e}^{-\alpha T}} \qquad, \hspace{0.5cm} I_\circ = \frac{g K_i\beta}{1-\mathrm{e}^{-\beta T}}
$$
are the magnitudes of the fields immediately after the synchronous spike emission. 
The constants $E_\circ$ and $I_\circ$ result self-consistently from the sum of the remaining field at the end of the period $T$ 
plus the contribution from the spike emission.

\begin{figure}
\center{\includegraphics[width=0.43\textwidth]{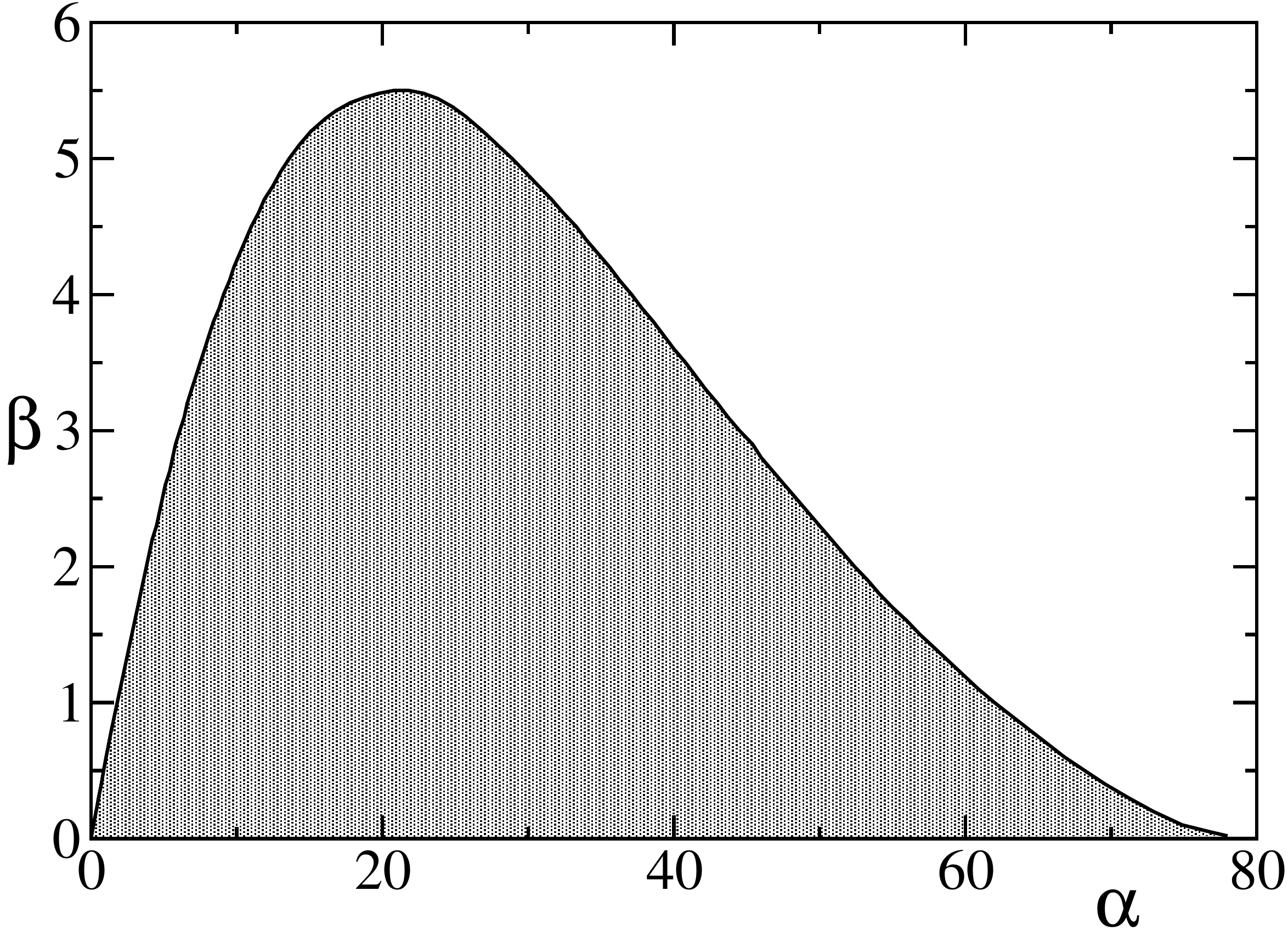}}
\caption{Phase diagram of synchronous regimes. Period-1 solutions exist above the curve in the white area. 
Crossing the curve, in the grey area, period-2 solutions first emerge followed (further down) by longer period
regimes.}
\label{fig:period}
\end{figure}

In the present paper, we focus on the stability of the synchronous period-1 solution (i.e. the initial configuration is 
exactly recovered after one spike emission).
For long inhibitory pulses (small inhibitory decay rate $\beta$) we observed also stable period-2 and higher order periodic solutions.  Fig.~\ref{fig:period} shows the transition from stable period-1 to period-2 solution in the $\alpha,\beta$ plane (from top to bottom). 
Higher order periodic solutions appear underneath that curve in the shaded area and result in a synchronous bursting dynamics.

\section{Linear stability analysis}\label{sec:linstab}

\subsection{General theory}

In this section we present the stability analysis of a synchronous state in the period-1 regime.
At variance with Ref.~\cite{Olmi14}, we do not construct the corresponding map, which means that the phase-space dimension is
not reduced by a suitable Poincar\'e section and the presence of a neutral direction is preserved. Actually this property can be
even used to double check the correctness of the final result. 

\begin{figure}
\center{\includegraphics[width=0.48\textwidth]{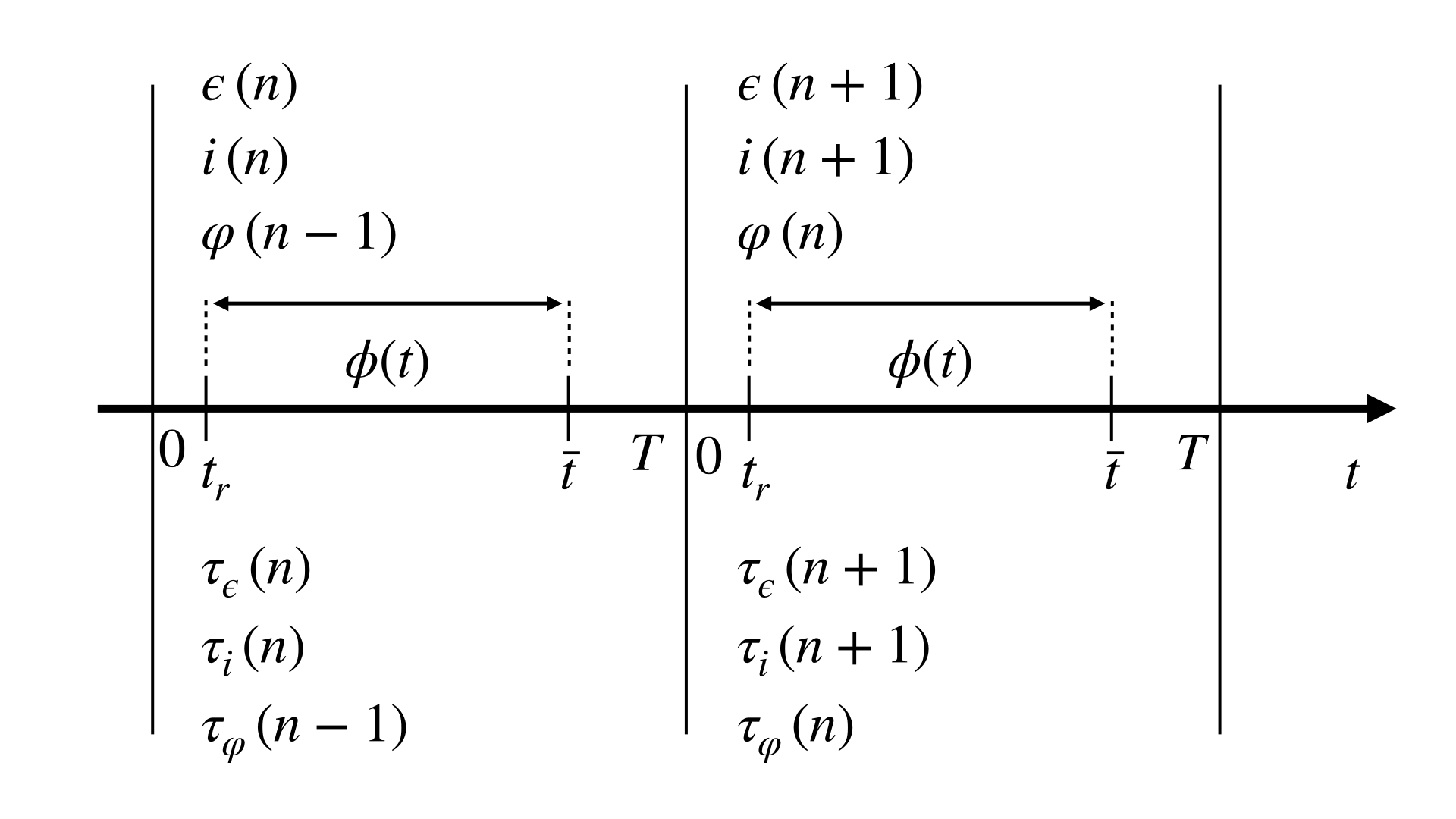}}
\caption{An illustration of the perturbation analysis in time $t$ for the synchronous state.}
\label{fig:timescheme}
\end{figure}

We start introducing a stroboscopic representation and focus on a weakly perturbed synchronous configuration
\begin{eqnarray*}
E^j(n) &=& E_r + \epsilon^j(n)  \\
I^j(n) &=& I_r + i^j(n) \\
\Phi^j(n) &=& \Phi_r + \varphi^j (n)
\end{eqnarray*}
where all variables are determined at the end of consecutive refractory periods. As shown in
Fig.~\ref{fig:timescheme} and clarified in the following, it is convenient to refer $\varphi^j(n)$ to one period
later with respect to $\epsilon^j(n)$ and $i^j(n)$.
The fields $E_r=E_\circ \mathrm{e}^{-\alpha t_r}$, $I_r=I_\circ \mathrm{e}^{-\beta t_r}$, and
$\Phi_r =0$ do not depend on $n$, as the reference trajectory is periodic of period $T$.
The overall perturbation can be represented as a $3N$ dimensional vector $[\boldsymbol{\epsilon}(n),\boldsymbol{i}(n),\boldsymbol{\varphi}(n)]$.
For the future sake of simplicity, it is convenient to introduce also a second representation in terms of time shifts,
$\boldsymbol{v}(n)= [\boldsymbol{\tau}_\epsilon(n),\boldsymbol{\tau}_i(n),\boldsymbol{\tau}_\varphi(n)]$, where
\begin{eqnarray}
\boldsymbol{\tau}_\epsilon(n)  &=&  \boldsymbol{\epsilon}(n)/\dot E_r \nonumber \\
\boldsymbol{\tau}_i(n) &=& \boldsymbol{i}(n)/\dot I_r \quad , \label{eq:newrep} \\
\boldsymbol{\tau}_\varphi(n) &=& \boldsymbol{\varphi}(n)/\dot \Phi_r \nonumber
\end{eqnarray}
and $\dot E_r$, $\dot I_r$ and $\dot \Phi_r$ all denote time derivates at the end of a refractory period.
In practice $\boldsymbol{\tau}_x$ corresponds to the time shift of the original trajectory to match the current perturbed state.
The recursive transformation can be formally written as
\begin{equation}\label{eq:lingen}
    \boldsymbol{v}(n+1) = \mathds{L}{\boldsymbol{v}}(n)  \; .
\end{equation}
Our next task is to determine the operator $\mathds{L}$. We start from the evolution equation of the excitatory field,
\begin{equation}\label{eq:efieldp}
E^j(n+1)=\mathrm{e}^{-\alpha T} E^j(n) + \alpha \sum_k G_{j,k} P_{k,k} \mathrm{e}^{-\alpha t^k(n)},
\end{equation}
where $t^k(n)$ is the time elapsed since the arrival of the spike sent by the $k$th neuron in the $n$th iterate.

Since the trajectory is close to the synchronous periodic orbit, 
${E}^j(n+1)=E_r + {\epsilon}^j(n+1)$, and $t^k(n)=t_r + \tau_\varphi^k(n)$.
Up to first order in the perturbations, Eq.~(\ref{eq:efieldp}) yields,
\begin{equation}\label{eq:lineps}
\epsilon^j(n+1) = \mathrm{e}^{-\alpha T}\epsilon^j(n) - \alpha^2 \mathrm{e}^{-\alpha t_r} \sum_{k}G_{j,k}P_{k,k} \tau_\varphi^k(n) \; ,
\end{equation}
or, in vector notations,
\begin{equation}\label{eq:lineexc}
\boldsymbol{\epsilon}(n+1)= A_e\boldsymbol{\epsilon}(n) - C_e\mathbf{GP} \boldsymbol{\tau}_\varphi(n)
\end{equation}
where
\[
A_e =\mathrm{e}^{-\alpha T} \qquad , \hspace{0.3cm} C_e=\alpha^2 \mathrm{e}^{-\alpha t_r}.   
\]
A similar analysis for the inhibitory field leads to
\begin{equation}\label{eq:lineinib}
\boldsymbol{i}(n+1) = A_i\boldsymbol{i}(n) - C_i\mathbf{G}\left(\mathds{1}-\mathbf{P}\right)\boldsymbol{\tau}_\varphi(n)
\end{equation}
where $\mathds{1}$ is the $N\times N$ identity matrix, while
\[
A_i =\mathrm{e}^{-\beta T} \qquad , \hspace{0.3cm} C_i=g \beta^2 \mathrm{e}^{-\beta t_r}.   
\]
Notice that, at variance with the previous case, there is an extra factor $g$ in the definition of $C_i$ to account for
the larger amplitude of the inhibitory spikes.

Finally, we deal with phase dynamics. The core of the transformation is the mapping between the amplitude 
$\boldsymbol{\varphi}(n)$ of the perturbation at time $t_r$ and the amplitude
$\bar{\boldsymbol{\varphi}}(n+1)$ at time $\overline t$, which can be formally written as
\begin{equation}\label{eq:linphase}
\bar{\boldsymbol{\varphi}}(n+1) = S_e \boldsymbol{\epsilon}(n+1) + S_i \boldsymbol{i}(n+1)  + S_\phi \boldsymbol{\varphi}(n) \; .
\end{equation}
This transformation is diagonal (it is the same for all components); the three unknown
parameters, $S_e$, $S_i$, and $S_\phi$, can be determined by integrating the equation obtained from the
linearization of Eq.~(\ref{eq:mod1}). To separate the notation of the stroboscopic phase perturbation $\varphi(n)$ from the continuously developing phase perturbation between $t_r$ and $\overline{t}$ we introduce $\phi(t)$ ,
\begin{equation}\label{7}
\dot{\phi} = J \, \Gamma'(\Phi) (E(t)-I(t))\phi
  + J \, \Gamma(\Phi)\left [\mathrm{e}^{-\alpha(t-t_r)}\epsilon - \mathrm{e}^{-\beta(t-t_r)}i \right ].
\end{equation}
Upon setting $[\epsilon,i,\phi(t_r)] = [1,0,0]$, $[0,1,0]$, and $[0,0,1]$,
$\phi(\overline t)$ corresponds to $S_e$, $S_i$, and $S_\phi$, respectively.
Once $\bar{\boldsymbol{\varphi}}(n+1)$ is known from Eq.~(\ref{eq:linphase}), it can be transformed into
the corresponding time shift
\[
\boldsymbol{\tau}_\varphi(n+1) = \frac{\bar{\boldsymbol{\varphi}}(n+1)}{\dot {\overline\Phi} }
\]
where 
\[
\dot {\overline \Phi}  = 1+ J \, \Gamma(\overline \Phi) \left( E_{\circ}\mathrm{e}^{-\alpha\overline{t}} - I_{\circ}\mathrm{e}^{-\beta\overline{t}}\right) 
\] 
is the time derivative of the phase in the point where the neuron stops feeling the action of the field.
In between $\overline t$ and $T$, the oscillators evolve with the same velocity and no adjustment of the
time shift can be expected.
The transformation is completed by Eq.~(\ref{eq:newrep}), which allows mapping $\boldsymbol{\varphi}(n)$ onto the corresponding
time shift and obtaining 
\begin{equation}\label{eq:lintau}
\dot {\overline \Phi} \ \boldsymbol{\tau}_\varphi(n+1) = S_e \boldsymbol{\epsilon}(n+1) + 
     S_i \boldsymbol{i}(n+1)  + S_\phi \dot \Phi_r \boldsymbol{\tau}_\varphi(n) \; .
\end{equation}
With the help of Eqs.~(\ref{eq:lineexc},\ref{eq:lineinib}), we find 
\begin{eqnarray}\label{eq:lintau3}
\dot {\overline \Phi} \ \boldsymbol{\tau}_\varphi&&(n+1) = 
A_e S_e \boldsymbol{\epsilon}(n) +
A_i S_i \boldsymbol{i}(n) - \\
&& \left[
C_e S_e \mathbf{GP} + C_i S_i \mathbf{G}\left(\mathds{1}-\mathbf{P}\right) - S_\phi \dot \Phi_r 
\right ]\boldsymbol{\tau}_\varphi(n) \; .  \nonumber
\end{eqnarray}
or, in a more compact form,
\begin{equation}\label{eq:lintau4}
\dot {\overline \Phi} \ \boldsymbol{\tau}_\varphi(n+1) = 
A_e S_e \boldsymbol{\epsilon}(n) + A_i S_i \boldsymbol{i}(n) -
\boldsymbol{M} \boldsymbol{\tau}_\varphi(n) \; .
\end{equation}
where $\mathbf{M}$ is an  $N\times N$ matrix whose entries are defined as follows,
\begin{align*}
& \mathbf{M}_{jk} = \begin{cases}
C_e S_e ,\hspace{0.2cm}&\text{if $k\rightarrow{j}$, $k\leq N_E$}\\
C_i S_i ,\hspace{0.2cm}&\text{if $k\rightarrow{j}$, $N_E< k\leq N$}\\
-S_\phi \dot \Phi_r ,\hspace{0.2cm}&\text{if $j=k$}\\
0,\hspace{0.2cm}&\text{no connection from $k$ to $j\ne k$}.
\end{cases}
\end{align*}
For homogeneity reasons, it is convenient to express all of the three recursive relations in terms of the components of the $\boldsymbol{v}$ vector,
\begin{eqnarray}
\boldsymbol{\tau}_\epsilon (n+1) &=& A_e\boldsymbol{\tau}_\epsilon(n) - \frac{C_e}{\dot E_r}\mathbf{GP} \boldsymbol{\tau}_\varphi(n) \nonumber \\
\boldsymbol{\tau}_i(n+1) &=& A_i\boldsymbol{\tau}_i(n) - \frac{C_i}{\dot I_r}\mathbf{G}\left(\mathds{1}-\mathbf{P}\right)\boldsymbol{\tau}_\varphi(n) \label{eq:lintot} \\
\boldsymbol{\tau}_\varphi(n+1) &=&
A_e S_e \frac{\dot E_r}{\dot {\overline \Phi}} \boldsymbol{\tau}_\epsilon(n) +
A_i S_i \frac{\dot I_r}{\dot {\overline \Phi}} \boldsymbol{\tau}_i(n) -
\frac{\boldsymbol{M} \boldsymbol{\tau}_\varphi(n)}{\dot {\overline \Phi}} \; .\nonumber
\end{eqnarray}
Now let us consider a homogeneous perturbation, such that $\boldsymbol{\tau}_\epsilon = \boldsymbol{\tau}_i =\boldsymbol{\tau}_\varphi$.
This perturbation must be mapped exactly onto itself, since it corresponds to a time shift of the whole orbit.
Let us see what this amounts to.
From the first of the above equations, we have that
\[
1 = A_e - C_e K_e/\dot E_r  \; .
\]
By looking at the definition of the various quantities, we can see that the equality is indeed satisfied. This is 
because $C_e /\dot E_r = - (1-A_e)/K_e$. Analogously, we can verify that $C_i /\dot I_r = - (1-A_i)/K_i$,
so that we can rewrite the transformation as
\begin{eqnarray}
\boldsymbol{\tau}_\epsilon (n+1) &=& A_e\boldsymbol{\tau}_\epsilon(n) +  \boldsymbol{0} \boldsymbol{\tau}_i(n)  +
\frac{1-A_e}{K_e}\mathbf{GP} \boldsymbol{\tau}_\varphi(n) \nonumber \\
\boldsymbol{\tau}_i(n+1) &=& \boldsymbol{0}\boldsymbol{\tau}_\epsilon(n)+ A_i\boldsymbol{\tau}_i(n) + \frac{1-A_i}{K_i}\mathbf{G}\left(\mathds{1}-\mathbf{P}\right)\boldsymbol{\tau}_\varphi(n) \nonumber \\
\boldsymbol{\tau}_\varphi(n+1) &=&
B_e \boldsymbol{\tau}_\epsilon(n) + B_i \boldsymbol{\tau}_i(n) -
\frac{\boldsymbol{M} \boldsymbol{\tau}_\varphi(n)}{\dot {\overline \Phi}}\label{eq:lintot2}
\end{eqnarray}
where $B_e = A_e S_e \dot E_r/{\dot {\overline \Phi}}$ and
$B_i = A_i S_i \dot I_r/{\dot {\overline \Phi}}$.

By playing the same game of homogenous perturbations with the last equation of Eq.~(\ref{eq:lintot}), we find that
\[
{\dot {\overline \Phi}} =  \dot E_rS_e + \dot I_rS_i + \dot \Phi_r S_\phi \; .
\]
Direct numerical simulations confirm that this condition is satisfied, as it should, since it implies that a homogeneous 
shift of the phase of all oscillators is time invariant. 

Altogether Eq.~(\ref{eq:lintot2}) is a representation of the linear operator $\mathds{L}$ formally introduced in 
Eq.~(\ref{eq:lingen}). The eigenvalues of $\mathds{L}$ are the so-called Floquet multipliers $Z_i$; the synchronous
solution is stable if the modulus of all multipliers is smaller than 1\footnote{With the exception of the unit multiplier
associated with a time shift of the trajectory.}.
One can equivalently refer to the Floquet exponents $\lambda_i = \log |Z_i|$ that we also call Lyapunov exponents with
a slight stretch of the notations.

For $\alpha,\beta \gg 1$ the fields are exponentially small when the neurons reach the threshold. 
In this limit, the fields behave as {\it slaved} variables and their contribution can be neglected in the stability
analysis, which reduces to diagonalizing an $N\times N$ matrix,
\begin{equation}
\boldsymbol{\tau}_\varphi(n+1) = 
- \boldsymbol{M} \boldsymbol{\tau}_\varphi(n) \; ,\label{eq:lintot3}
\end{equation}
(notice that $\dot {\overline \Phi}$ can be safely set equal to 1, as the coupling is negligible at time
$\overline t$).

\subsection{Transversal Lyapunov exponent}

A simpler approach to assess the stability of the synchronous regime consists in investigating the stability of a single
neuron subject to the external periodic modulation resulting from the network activity.
The corresponding growth rate $\lambda_c$ of infinitesimal perturbations  is called transversal or 
conditional Lyapunov exponent. In mean-field models, this approach leads to the same result obtained by implementing 
a more rigorous theory which takes into account mutual coupling.
Let the time shift at the end of a refractory period be equal to $\tau_r$; the corresponding phase shift is therefore
\begin{equation}\label{eq:transv0}
\phi(t_r) = \dot \Phi(t_r)\tau_r = \{1+ J \Gamma(0) [E(t_r)-I(t_r)]\}\tau_r \; .
\end{equation}
From time $t_r$ up to time $\overline{t}$ the phase shift evolves according to 
simplified version of Eq.~(\ref{7}), 
\begin{equation}\label{eq:transv}
\dot{\phi} = \Gamma'(\Phi) (E(t)-I(t))\phi \; ,
\end{equation}
where we have neglected the variation of field dynamics, since the field is treated as an external forcing.
As a result,
\begin{equation}\label{eq:tr1}
    \phi(\overline{t}) = \mathrm{e}^D \phi(t_{r}) \; ,
\end{equation}
where, with reference to the PRC Eq.~(\ref{eq:PRC}), 
\begin{equation}\label{eq:ddef}
D = \frac{E_\circ}{\beta} \left[\mathrm{e}^{-\beta \overline{t}}-\mathrm{e}^{-\beta t_{r}}\right]-
\frac{I_\circ}{\alpha}\left[\mathrm{e}^{-\alpha \overline{t}}-\mathrm{e}^{-\alpha t_{r}}\right]
\end{equation}
The corresponding time shift is
\[
\overline{\tau} = \frac{\phi(\overline{t})}{\dot {\overline\Phi} }
\]
The shift $\overline{\tau}$ carries over unchanged until first the threshold $\phi=1$ is crossed and then 
the new refractory period ends. Accordingly, from Eqs.~(\ref{eq:transv0},\ref{eq:tr1}),
the expansion $R$ of the time shift over one period (a sort of Floquet multiplier) can be written as

\begin{equation}\label{eq:R}
R = \frac{\overline{\tau}}{\tau_r} = 
\frac{1+ J \Gamma(0) [E(t_r)-I(t_r)]}{\dot {\overline \Phi}} \mathrm{e}^D
\end{equation}
This formula is substantially equivalent to Eq.~(54) of Ref.~\cite{Olmi14} ($\Lambda_{ii}$ corresponds to $R$),
obtained while studying a single population under the action of $\alpha$-pulses. 
An additional marginal difference is that while in Ref.~\cite{Olmi14} the single neuron dynamics is described by a non uniform
velocity field $F(x)$ and homogeneous coupling strength, here we refer to a constant velocity and a phase-dependent
PRC, $\Gamma(\phi)$. 

The corresponding conditional Lyapunov exponent is
\begin{equation}\label{eq:lyapcond}
\lambda_c = \frac{\ln |R| }{T}=\frac{D+\ln{{\left|[1+J\Gamma(0)(E(t_{r})-I(t_r))]/{\dot {\overline \Phi}}\right|}}}{T}.
\end{equation}
It is the sum of two contributions: the former one accounting for the linear stability 
of the phase evolution from reset to threshold ($D/T$); the latter term arises from the different velocity
(frequency) exhibited at threshold and at the end of the refractory period.
Notice the in the limit of short pulses, the field amplitude at time $\overline t$ can be set equal to zero, thereby
neglecting the corresponding exponential terms in Eq.~(\ref{eq:ddef}) and assuming $\dot {\overline \Phi}= 0$.

\section{Application}\label{sec:app}

We now implement the general formalism in the case of the PRC defined by Eq.~(\ref{eq:PRC}), considering a network with
$N=1000$ neurons, a 10\% connectivity (i.e. $K=100$ with $K_e=80$ and $K_i=20$), and $g=5$; the coupling strength is 
assumed to be $J=0.03$, while the refractory time is $t_r=0.03$. This setup, characterized by a slight prevalence of
inhibition ($g K_i \gtrsim K_e$), is often adopted in the study of balanced regimes (see e.g.~\cite{ostojic_2014nat}).

\begin{figure}
\center{\includegraphics[width=0.43\textwidth]{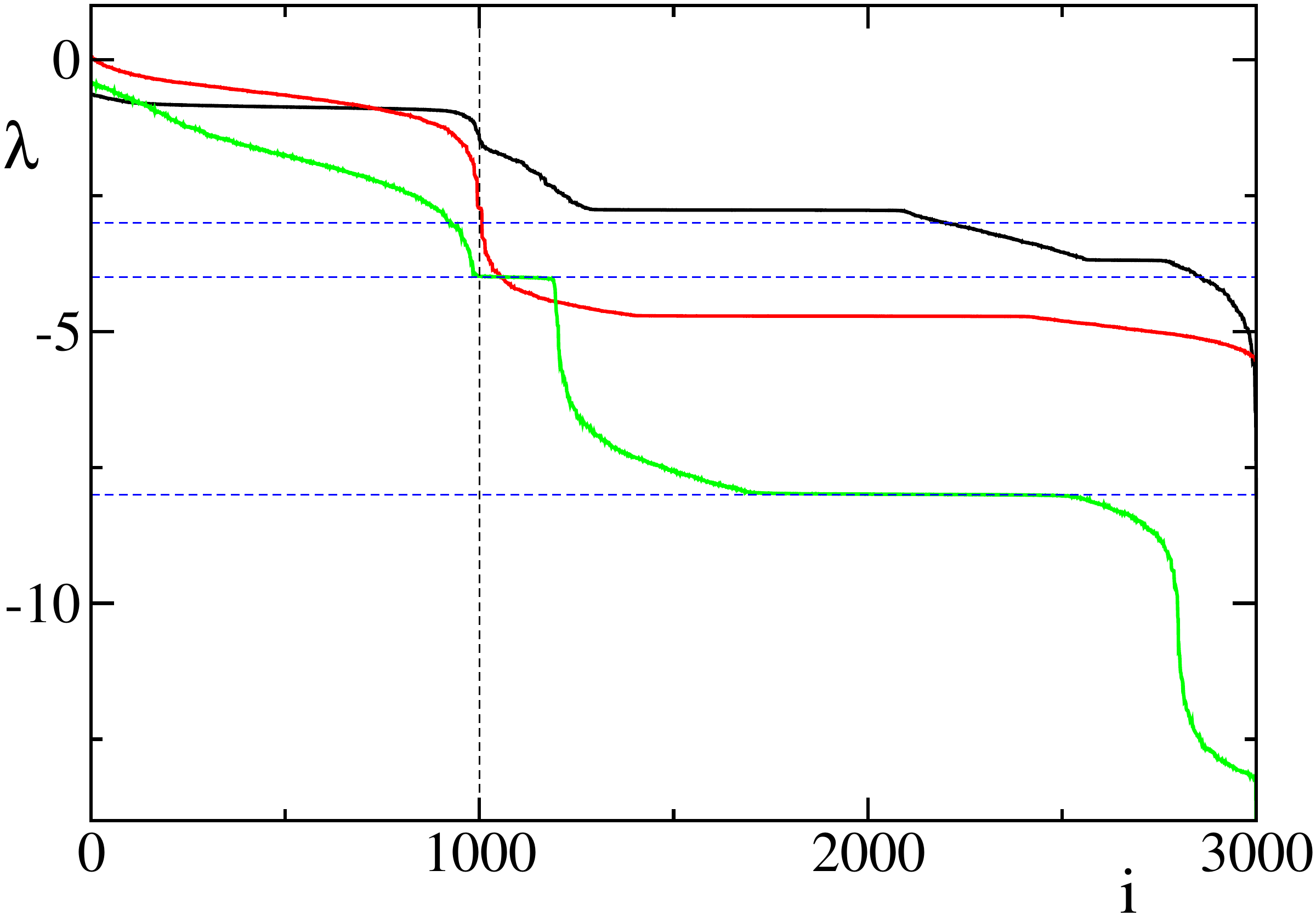}}
\caption{The total Lyapunov spectrum, for $(\alpha,\beta)$ equal to (4,3), (4,4), and (4,8) -- top to bottom;
the coupling strength is $J=0.03$, as for all of our simulations, while $N=1000$. The three horizontal dashed lines correspond
to the three different rates used to identify $\alpha$ and $\beta$ values. The vertical dashed line separates the part of the
spectrum which, for large $\alpha$ and $\beta$ values, is related to the actual network structure.}
\label{fig:lyap_spec}
\end{figure}

The resulting Floquet spectra are presented in Fig.~\ref{fig:lyap_spec} for three different pairs of not-too-large $\alpha$ and
$\beta$ values.
Rather than diagonalizing the matrix defined by Eq.~(\ref{eq:lintot2}), the 3,000 exponents have been determined by 
implementing a standard algorithm for the computation of Lyapunov exponents \cite{PikovskyPoliti}. 
The larger are $\alpha$ and $\beta$, the more step-like is the spectral shape, the two lower steps being located around the 
decay rate (i.e. the inverse pulse-width) of the pulses (see the three horizontal dashed lines, which correspond to 
$\lambda= -3$, -4, and -8, respectively). This is sort of expected, since the field dynamics basically amounts to 
a relaxation process controlled by the respective decay rate. 
Anyhow, since the overall stability is determined by the largest exponents, it is sufficient to restrict the analysis 
to the first part of the spectrum (to the left of the vertical dashed line in Fig.~\ref{fig:lyap_spec}), which, in the 
limit of large $\alpha$ and $\beta$, can be directly determined by diagonalizing the matrix defined in Eq.~(\ref{eq:lintot3}).

\begin{figure}
\center{\includegraphics[width=0.45\textwidth]{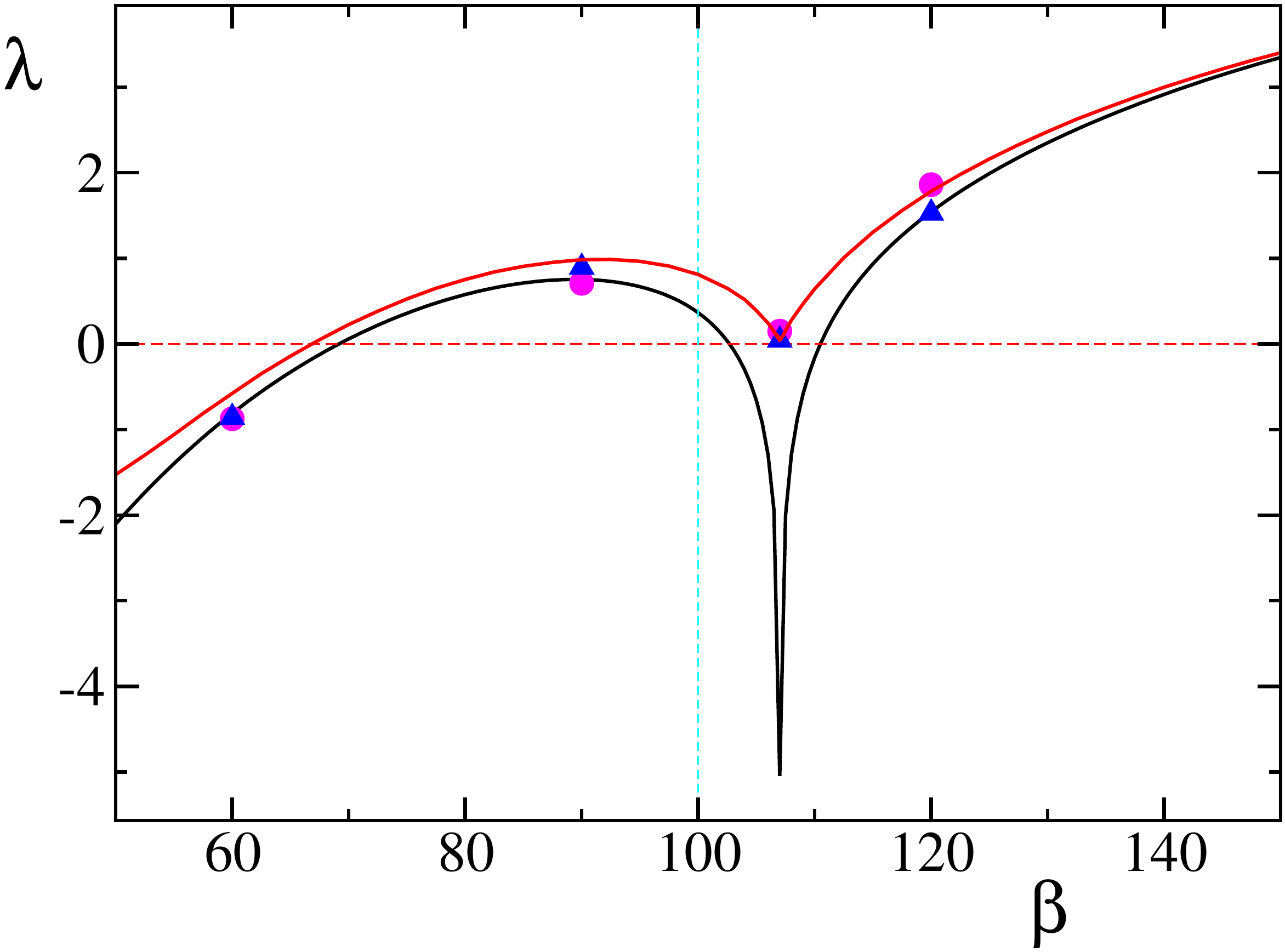}}
\caption{The maximal Lyapunov/Floquet exponent $\lambda_M$ (upper red curve) vs. $\beta$ for a network with  $N=10,000$, 
$\alpha=100$ and the other parameters set as in the previous figure.
The black curve corresponds to the transversal/conditional exponent $\lambda_c$, while full dots and triangles result 
from the computation of the finite-amplitude Lyapunov exponent $\lambda_f$ for $\sigma = 10^{-2}$ and $10^{-3}$, respectively.}
\label{fig:short_pulses}
\end{figure}

The dependence of the maximum exponent $\lambda_M$ on the (inverse) pulse-width of the inhibitory spikes is reported in 
Fig.~\ref{fig:short_pulses} (see the upper red curve). In this case, the Floquet exponent has been obtained by diagonalizing 
the matrix in Eq.~(\ref{eq:lintot3}) for a system size $N=10,000$ and a connectivity $K=1000$ ($K_e=800$, $K_i=200$).

The vertical dashed line corresponds to the symmetric case, where both excitatory and inhibitory neurons have the same width (and 
shape). Interestingly, the stability, determined by the largest non zero exponent, (the always present $\lambda=0$, corresponds
to the neutral stability associated to a time shift of the trajectory) depends strongly on the relative excitatory/inhibitory
pulse width and can even change sign: the synchronous solution is stable below 
$\beta=67$~\footnote{For the sake of completeness notice that by further decreasing $\beta$, the stability changes again.}.
Additionally, there is evidence of a sort of singularity around $\beta=107$, when the inhibitory spikes are
slightly shorter than the excitatory ones.

Given the finite dimension of the matrices, sample-to-sample fluctuations are expected.
Such fluctuations are, however, rather small, as testified by the smoothness of the red curve in Fig.~\ref{fig:short_pulses}.
In fact, the single values of the Floquet exponents have been obtained not only by varying the (inhibitory) pulse-width, 
but also considering different network realizations. Although small, the fluctuations prevent drawing definite 
conclusions about the singularity seemingly displayed by the derivative of $\lambda_M(\beta)$ around $\beta=107$.

In the limit of a fully connected network, we expect a perfectly degenerate spectrum (all directions are mutually equivalent)
and $\lambda_M$ equal to the conditional Lyapunov exponent $\lambda_c$ defined in Eq.~(\ref{eq:lyapcond}). 
The lower black curve reported in Fig.~\ref{fig:short_pulses} corresponds to $\lambda_c$; except for a narrow region
around $\beta=107$, $\lambda_c$ is always close to (lower than) $\lambda_M$. This means that the mean-field approximation
still works pretty well in a network of 10,000 neurons with a 10\% connectivity.

\begin{figure}
\center{\includegraphics[width=0.45\textwidth]{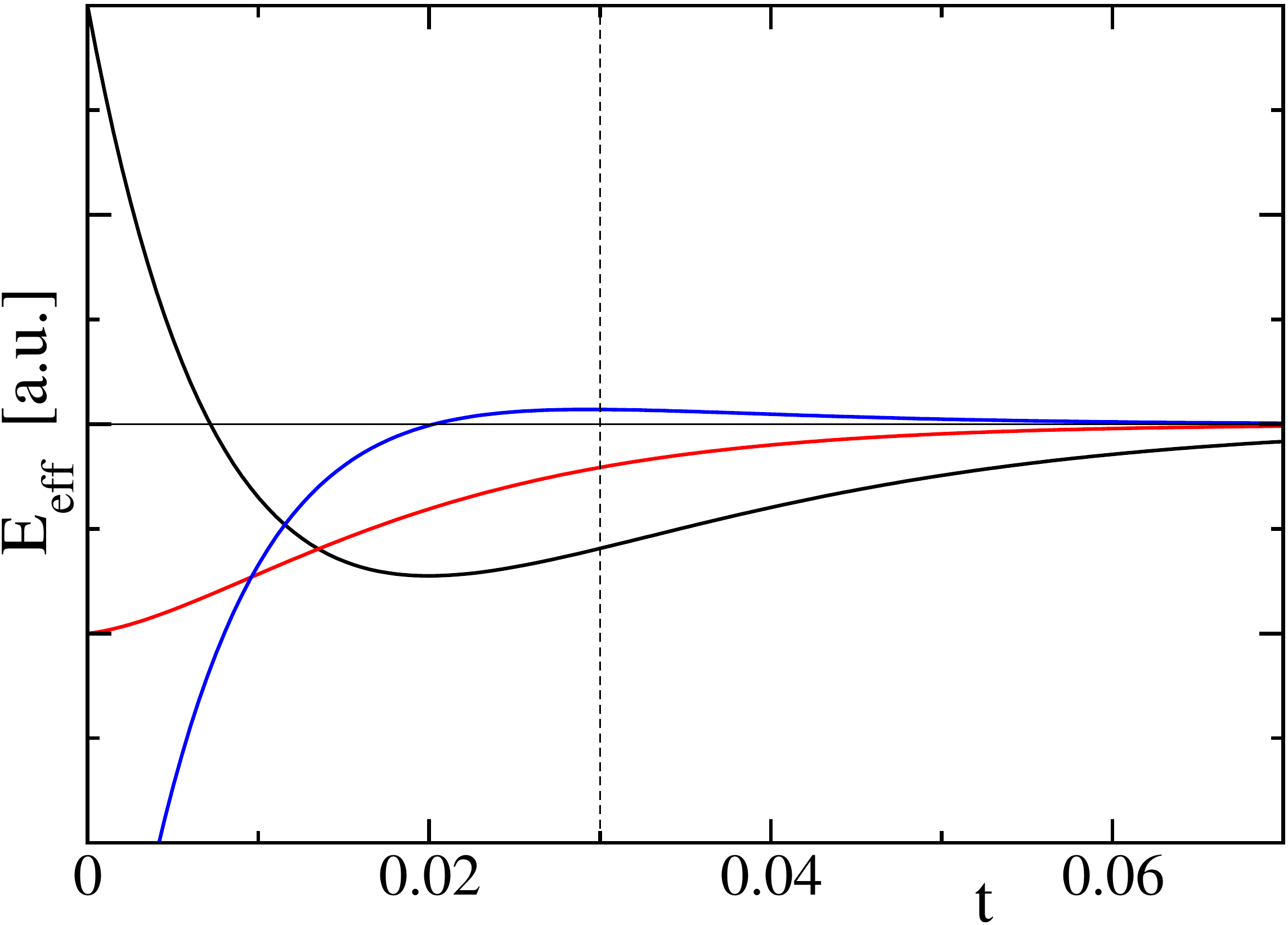}}
\caption{Effective field shape for $\alpha=100$ and $\beta=60$ (black), $\beta=90$ (red), and $\beta=120$ (blue). The vertical
dashed line identifies the end of the refractory period.}
\label{fig:pulses}
\end{figure}

The explicit formula Eq.~(\ref{eq:lyapcond}) helps also to shed light on the $\beta$ dependence of the network stability.
The main responsible for the qualitative changes observed around $\beta=107$ is the logarithmic term, arising from the
difference between the velocity at threshold (equal to 1, irrespective of the $\beta$-value) and the velocity at the
end of the refractory period. This latter velocity is determined by the effective field
$E_{eff}(t_r)= E(t_r)-I(t_r)$ which in turn strongly depends on the relative pulse-width. The time dependence of $E_{eff}$ can be
appreciated in Fig.~\ref{fig:pulses}, where we report the trace for three different $\beta$ values (60, 90, and 120)
and the same $\alpha = 100$.
There, we see that even the sign of the effective field may change;
for $\beta=120$, $E_{eff}$ is initially negative because inhibition dominates, but above $t=0.02< t_r$ the slower decay
of the excitatory pulses takes over, so that the effective field amplitude is positive at the end of refractoriness.
For $\beta=90< \alpha=100$, inhibition prevails at all times and the effective field is thereby negative for $t=t_r$.
Finally, for $\beta=60$, excitation initially prevails, but inhibition takes soon over. 

From Eq.~(\ref{eq:lyapcond}), we see that the sign of the logarithmic contribution changes depending whether the argument 
is smaller or larger than 1. More precisely if the effective field is negative but larger than $-2/(J\Gamma(0))$,
the discontinuity of the velocity tends to stabilize the synchronous regime; 
if $E_{eff}(t_R)= -1/J\Gamma(0)$ the orbit is even superstable, i.e. the Lyapunov exponent is infinitely negative.
This is precisely what happens for $\beta \approx 107$.
Altogether, the $\beta$ interval around $107$ separates the region where the expansion/contraction factor 
is positive (to the right), from the region where it is negative (to the left).

The sign of the multiplier has a simple explanation:
$1+J \, \Gamma(0)  \, E_{eff}(t_r) < 0$ means that the phase velocity is negative at the of the refractory period. 
Therefore, if one follows two nearby neurons -- one leading over the other before reaching the threshold --
then at the end of refractoriness, the leading neuron becomes the lagging one, as they initially move in the 
``wrong" direction\footnote{Later, the velocity changes sign becoming positive, but this does not modify the ordering.}. 
This explains how the pulse-width may affect the stability.

\begin{figure}
\center{\includegraphics[width=0.45\textwidth]{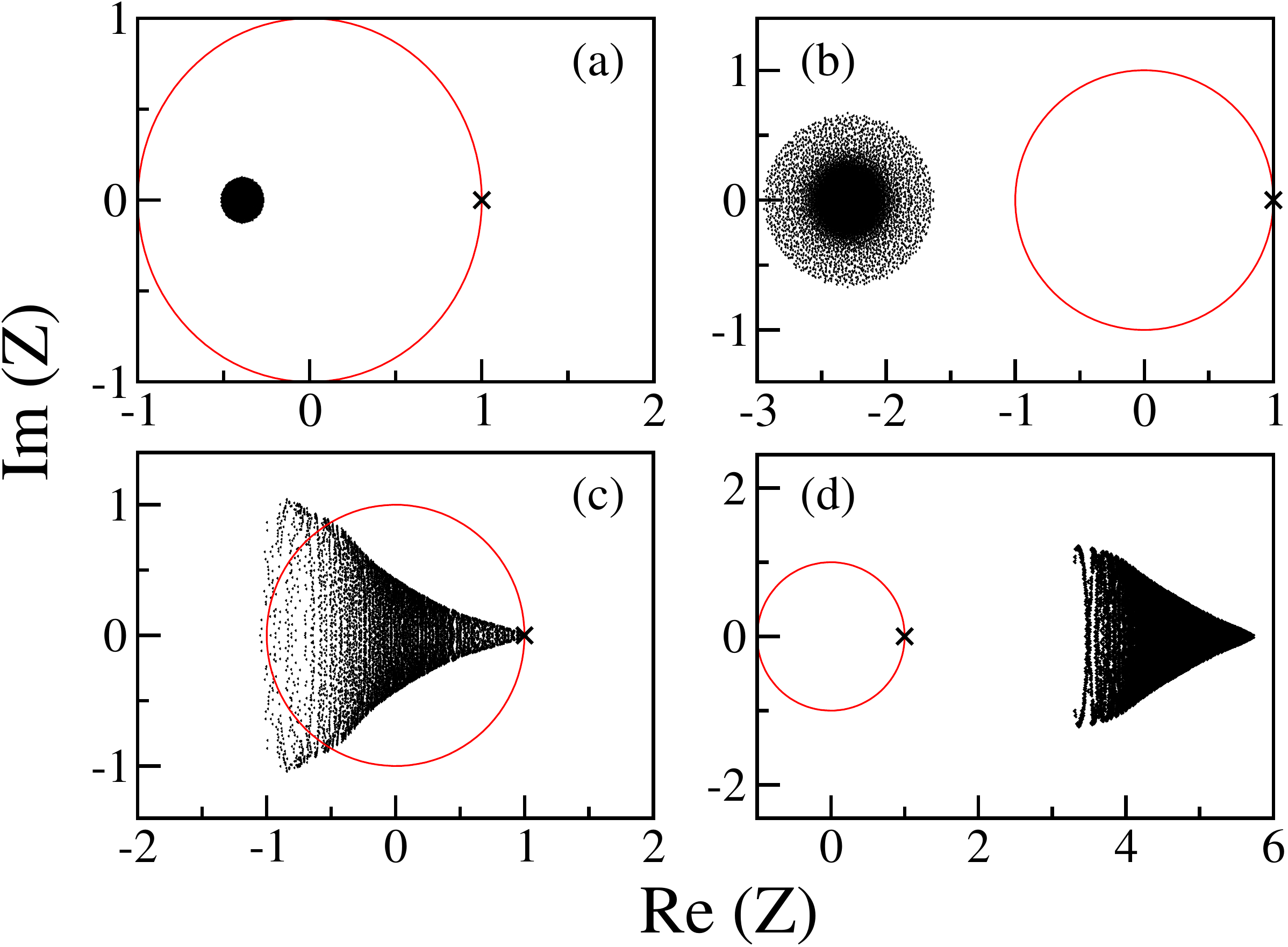}}
\caption{The distribution of the complex eigenvalues (black dots) for short pulses, $N=10,000$, $\alpha=100$, and $J=0.03$. The 
red curve highlights the unit circle. The black cross at $(1,0)$ singles out the always present eigenvalue associated with the periodic motion. The four panels represent the results for different decay rates of the inhibitory pulses $\beta$. In particular $\beta = 60$, $90$, $107$ and $120$ for panel (a) - (d), respectively.}
\label{fig:eigenval}
\end{figure}

So far we have referred to the Floquet exponents, without paying attention to the phase of the multipliers.
In Fig.~\ref{fig:eigenval} we report both real and imaginary part of all eigenvalues for four different $\beta$ values.

For $\beta = 60$ and 90, the eigenvalues (except for $Z=(1,0)$) are distributed within a circle (see panels a and b).
This is reminiscent of Girko's theorem~\cite{girko}, which states that the eigenvalues of an $N \times N$ random matrix 
with independent and identically distributed entries (with zero mean zero and variance equal to $1/N$), are 
uniformly distributed over the unit disc. However, it is not obvious how to adapt/extend this theorem to the present
context, since the
matrix $\bf M$ although being random does not satisfy several of the required assumptions, starting from the off-diagonal elements which
take only three different values and their average is non zero.

Returning to Fig.~\ref{fig:eigenval}, for $\beta=60$ all the eigenvalues lie within the unit circle,
meaning that the synchronous solution is stable, while for $\beta=90$ {\it all} eigenvalues lie outside,
meaning it is fully unstable: any perturbation is amplified!

Above $\beta=100$, the spectrum changes shape, becoming funnel-like: for $\beta=120$ (panel d), all eigenvalues sit again 
outside the unit circle, meaning that the synchronous solution is fully unstable.
Interestingly, for $\beta=107$ (panel c), the funnel is almost entirely contained inside the unit circle, so that the resulting
(weak) instability is due to few complex eigenvalues lying on the upper-left and lower-left corners of the funnel.
As an additional remark, we can see that the eigenvalue with largest modulus (i.e. the one determining the stability) 
is real and negative for $\beta=60$ and $90$, while it is real and positive for $\beta=120$\footnote{In practice, depending
on the network realization, the leading eigenvalues may have a small imaginary component.}.
This is coherent with the behavior of the sign of the multiplier $R$ (see Eq.~(\ref{eq:R})), which changes
from positive to negative, while decreasing $\beta$. The qualitative differences observed in the region around
$\beta=107$ suggest that the ``singular" behavior exhibited by $\lambda_M$ is the signature of a true transition associated
with a change of the spectral structure.

Finally, a few words about the leading eigenvector. It must possess some special features which are responsible for its 
larger expansion rate. However, we have not found any correlation with obvious indicators such as an anomalously large 
outgoing connectivity. We have only observed that the vector components are distributed in a Gaussian way with zero
average.

\subsection{Finite-amplitude perturbations}

Finally, we have directly investigated the stability of the synchronous regime, by studying the evolution of small but 
finite perturbations under the action of the model Eqs.~(\ref{eq:mod1}-\ref{eq:mod2}) in the limit of short pulses.
By following the same strategy developed in tangent space, the perturbation amplitude has been quantified
as the temporal shift at a specific moment.
We find it convenient to identify the {\it specific moment} with the threshold-passing time $t^L(n)$ of the last neuron (in the
$n$th period). Provided the perturbation is small enough, all neurons are still in the refractory period 
and their phase is equal to 0 when the time is taken. The temporal shift of the $j$th neuron can be defined as
$\delta_j = t^L(n) -t_j(n)$, where $t_j(n)$ is its $n$th passing time. The perturbation amplitude is finally defined
as the standard deviation $\sigma(n)$ of all temporal shifts.
Given an initial distribution with a fixed $\sigma(0)$, it is let evolve to determine its value once the new set of spiking 
times is over. The ratio $R_f = \sigma(1) / \sigma(0)$ represents the contraction or expansion factor over one period $T$. 
Afterwards the standard deviation is rescaled to the original value $\sigma(0)$ to avoid it becoming either too large to be
affected by nonlinear effects or too small to be undetectable.
We have found that $\sigma(0) = 10^{-3}$, or $10^{-2}$ suffices to ensure meaningful results.
The corresponding (finite amplitude) Lyapunov exponent $\lambda_f$ 
is finally obtained by iterating this procedure to let the perturbation converge along the most expanding direction 
and thereby computing $\lambda_f = \ln{|R_f|}/T$. We have found that 50 iterates suffice to let the transient die out.

A crucial point is the integration time step, if the model is evolved by implementing an Euler algorithm.
In fact, the time step must be much smaller than the separation between ocnsecutive spike-times, since they have to be well resolved.
We have verified that setting the Euler integration time step $\Delta t$ at least $100$ times smaller than $\sigma(0)$ ensures
a sufficient accuracy. The numerical results, plotted in Fig.~\ref{fig:short_pulses} for four different
$\beta$ values (see the symbols), indeed confirm the theoretical predictions.

\section{Conclusions and open problems}
\label{sec:conclusion}

In this paper, we have developed a formalism to assess the stability of synchronous regimes in sparse networks 
of two populations of oscillators coupled via finite-width pulses.
The problem is reduced to the determination of the spectral properties of a suitable class of sparse random matrices.
Interestingly, we find that the relative width of excitatory and inhibitory spikes plays a crucial role even in the 
limit of narrow spikes, up to the point that the stability may qualitatively change.
This confirms once more that the $\delta$-spike limit is  and it is necessary to include the spike width into 
the modelling of realistic neuronal networks.

Our analytical treatment has allowed constructing the stability matrix, but deriving an analytical solution of the spectral problem 
remains an open problem. The conditional Lyapunov exponent provides an approximate expression for the maximum Floquet exponent.
It is quite accurate in a broad range of pulse-widths but fails to predict the weak instability occurring when inhibitory 
pulses are slightly narrower than excitatory ones. For relatively wider inhibitory pulses, numerical simulations suggest
that it will be worth exploring the possibility to extend the circular law of random matrices to sparse matrices of the type
herewith derived.

While mean-field models are characterized by a degenerate spectrum (all directions being  equally stable), here the degeneracy is
lifted by the randomness associated with the sparse connectivity. It is therefore desirable to understand which features
make some directions so special as to be characterized by a minimal stability. This is probably related to the presence of
closed loops of connections among oscillators which sustain an anticipated or retarded firing activity. Further studies are
required.

\begin{acknowledgements}
Afifurrahman was supported by an LPDP Indonesia fellowship.
\end{acknowledgements}

\bibliographystyle{apsrev4-1}
\bibliography{stability.bib}
 
\end{document}